\documentclass{ws-procs9x6}            % default, citations in superscript
\begin{document}
\title{Charmed excited baryons with heavy-quark spin symmetry}

\author{L. Tolos $^*$}

\address{Institut f\"ur Theoretische Physik, University of Frankfurt, \\ Max-von-Laue-Str. 1, 60438 Frankfurt am Main, Germany \\
Frankfurt Institute for Advanced Studies, University of Frankfurt, \\ Ruth-Moufang-Str. 1, 60438 Frankfurt am Main, Germany \\
 Institute of Space Sciences (ICE, CSIC), \\ Campus UAB, Carrer de Can Magrans, 08193, Barcelona, Spain\\
Institut d'Estudis Espacials de Catalunya (IEEC), \\ 08034 Barcelona, Spain\\
$^*$E-mail: tolos@th.physik.uni-frankfurt.de}

\author{J. Nieves and R. Pavao}

\address{Instituto~de~F\'{\i}sica~Corpuscular~(centro~mixto~CSIC-UV), \\ Institutos~de~Investigaci\'on~de~Paterna, Aptdo.~22085,~46071,~Valencia,Spain}

\begin{abstract}
Recently five $\Omega_c$  excited states have been reported by the LHCb Collaboration, four of them corroborated by Belle. The Belle Collaboration has also discovered in 2017 one excited $\Xi_c$ with mass of 2930 MeV. In view of these recent detections, we analyze the possible molecular description of these states, using a unitarized baryon-meson model that incorporates both chiral and heavy-quark spin symmetries in a consistent manner.  We pay a special attention to the renormalization procedure, so as to determine the robustness of our predictions. Within our model, we generate dynamically several $\Omega_c$ and $\Xi_c$ states. We find that, at least, three of our $\Omega_c$ states can be identified with the experimental ones and have spin-parity $J=1/2^-$ or $J=3/2^-$. Moreover, we find a plausible molecular description of not only the $\Xi_c(2930)$ state, but also $\Xi_c(2790)$, $\Xi_c(2815)$ and $\Xi_c(2970)$, reported in the PDG.  Interestingly, we determine that $\Xi_c(2930)$ and $\Xi_c(2970)$ are heavy-quark spin partners with $J=1/2^-$ and $J=3/2^-$, respectively, while obtaining that $\Xi_c(2930)$, $\Omega_c(3090)$ or the $\Omega_c(3119)$, and $\Sigma_c(2800)$ would belong to the same SU(3) multiplet.

\end{abstract}

\keywords{unitarized baryon-meson model, charmed baryonic excited states, heavy-quark spin symmetry}

\bodymatter

\section{Introduction}
\label{sec:intro}

The study of charmed baryonic excited states has been a matter of high interest over the past years in view of several recent discoveries. The LHCb Collaboration has reported the existence of five $\Omega_c$ excited states in the $\Xi_c^+ K^-$ spectrum in $pp$ collisions \cite{Aaij:2017nav}, four of them confirmed by the Belle Collaboration \cite{Yelton:2017qxg}. Also, the $\Xi_c(2930)$ state has been determined by Belle \cite{Li:2017uvv} in its decay to $\Lambda_c^+ K^-$ from $B^- \rightarrow K^- \Lambda_c^+ \bar \Lambda_c^-$ . Other excited $\Xi_c$ states are reported around 3 GeV, such as $\Xi_c(2790)$, $\Xi_c(2815)$ and $\Xi_c(2970)$  \cite{Tanabashi:2018oca}.

The understanding of the nature of these newly discovered states have triggered an important theoretical effort within molecular baryon-meson models \cite{Montana:2017kjw,Debastiani:2017ewu,Wang:2017smo,Chen:2017xat,Yu:2018yxl}, and, in particular, within a model that fully respects heavy-quark spin symmetry (HQSS), which is a proper QCD symmetry in the limit of large quark masses \cite{GarciaRecio:2008dp,Gamermann:2010zz,Romanets:2012hm,GarciaRecio:2012db,Garcia-Recio:2013gaa,Tolos:2013gta,Garcia-Recio:2015jsa}. The starting point is a ${\rm SU(6)}_{\rm lsf} \times {\rm HQSS}$ extension of the Weinberg-Tomozawa (WT) interaction, with "lsf'' indicating the light quark-spin-flavor symmetry. In Ref.~\refcite{Romanets:2012hm} five $\Omega_c$ states were generated dynamically, three $J=1/2$ and two $J=3/2$ bound states,  stemming from the most attractive ${\rm  SU(6)}_{\rm lsf}\times$ HQSS representations, but with masses below the LHCb predictions. Also, within the same model, nine $\Xi_c$  were obtained, below or close in mass to the experimental $\Xi_c(2790)$ and $\Xi_c(2815)$ states, while being far below in mass from $\Xi_c(2930)$ and $\Xi_c(2970)$. 

 All these predictions of masses, however, depend on the adopted renormalization scheme (RS). In Refs.~(\refcite{Nieves:2017jjx,Nieves:2019jhp}) we have revisited the RS used in Ref.~\refcite{Romanets:2012hm}. We have payed a special attention to the flavor-symmetry content of the ${\rm SU(6)}_{\rm lsf} \times {\rm  HQSS}$ model in order to determine the possible HQSS partners and siblings among the experimental states while predicting new ones. In the present work we summarize the results of Refs.~(\refcite{Nieves:2017jjx,Nieves:2019jhp}) on the $\Omega_c$ and $\Xi_c$ sectors.

\section{Formalism}
\label{sec:formalism}

We solve the Bethe-Salpeter equation in the on-shell approximation for the scattering amplitude  ($T^J$)  for a total angular momentum $J$ as
\begin{equation}
\label{eq:LS}
T^J(s)=\frac{1}{1-V^J(s) G^J(s)} V^J(s),
\end{equation}
with $V^J$ being the $S$-wave baryon-meson potential resulting from the  ${\rm SU(6)}_{\rm lsf}\times {\rm HQSS}$ WT interaction between pseudoscalar or vector mesons and the low-lying $1/2^+$ or $3/2^+$ baryons. The diagonal $G^J(s)$ matrix contains the baryon-meson loop functions for each channel $i$, that is,
\begin{equation}
G_i(s)=\overline{G}_i(s)+G_i(s_{i+}) .
\label{eq:div}
\end{equation}
These are logarithmically ultraviolet (UV) divergent, and need to be renormalized. The finite part, $\overline{G}_i(s)$ is obtained from Ref.~\refcite{Nieves:2001wt}, whereas the divergent contribution, $G_i(s_{i+})$, can be renormalized, either by
one subtraction at certain scale ($\sqrt{s}=\mu$) %
\begin{eqnarray}
G_i^\mu(s) =\overline{G}_i(s) - \overline{G}_i(\mu^2) ,
\label{eq:relation}
\end{eqnarray}
or  using a sharp-cutoff regulator $\Lambda$ in momentum space, that is,
\begin{equation}
G^{\Lambda}_i(s) =\overline{G}_i(s) + G_i^{\Lambda}(s_{i+}). \label{eq:uvcut2}
\end{equation}

The excited $\Omega_c$ and $\Xi_c$ states are obtained as poles of the scattering amplitudes in each $J$ sector  (see Refs.~(\refcite{Romanets:2012hm,Nieves:2017jjx,Nieves:2019jhp}) for details). 

\section{Results}
\label{sec:results}

\subsection{$\Omega_c$ excited states}
\label{sec:omegac}

In Ref.~\refcite{Romanets:2012hm} five excited $\Omega_c$ states with $J=1/2^-$ and $J=3/2^-$  were found, with masses
below 3 GeV, thus, being difficult to identify them with the LHCb results. These states were obtained renormalizing with one-substraction at a certain scale. In order to study the dependence of our results on the RS and the possible identification with the experimental observations, we employ a common UV
cutoff for all baryon-meson loops within a reasonable range. In this way we  avoid any uncontrolled reduction of a loop and an arbitrary change of the subtraction constants.

First we need to follow the original $\Omega_c$ states in the complex energy plane as we modify our prescription from one-subtraction to a common UV cutoff renormalization. Hence, we change each loop function by
\begin{equation}
G_i(s) = \overline{G}_i(s)-(1-x) \overline{G}_i(\mu^2)+x G_i^{\Lambda}(s_{i+}),
\end{equation}
with $x$ moving adiabatically from $0$ to $1$.  In this manner and by varying the value of UV cutoff, we find that (probably) at least three of the experimental states can be identified with three of our $\Omega_c$ \cite{Nieves:2017jjx}. As an example, for $\Lambda=1090$ MeV, we identify the experimental $\Omega_c(3000)$ and $\Omega_c(3090)/\Omega_c(3119)$ with two of our $1/2^-$ states, and $\Omega_c(3050)$ with one of our $3/2^-$. Also, $\Omega_c(3000)$ and $\Omega_c(3050)$ would be members of the same  ${\rm SU(6)}_{\rm lsf}$ $\times$ HQSS multiplet \cite{Nieves:2017jjx}.

Other models have also determined the molecular nature of some of the experimental $\Omega_c$ states. In particular, in Ref.~\refcite{Montana:2017kjw} the $\Omega_c(3050)$ and $\Omega_c(3090)$ were identified as $1/2^-$ states. In Ref.~\refcite{Debastiani:2017ewu} the $\Omega_c(3050)$ and
$\Omega_c(3090)$ were also obtained as $1/2^-$ states, due to the use of the same interaction for $J=1/2$ as Ref.~\refcite{Montana:2017kjw}, whereas the experimental $\Omega_c(3119)$ was obtained with $3/2^-$. The difference between the predictions of our model and these schemes comes from a different RS as well as  different interactions for the channels involving $D$, $D^*$ and light vector mesons with baryons, that are not fixed by chiral or HQSS symmetries.  With regards to a broad structure around  3188 MeV determined by LHCb \cite{Aaij:2017nav}, our results differ from those of Ref.~\refcite{Wang:2017smo}, as we cannot make any identification, since this state would come from a less attractive ${\rm SU(6)}_{\rm lsf}\times$HQSS representation. Also the prediction of Ref.~\refcite{Chen:2017xat}  of a loosely bound molecule of 3140 MeV is not confirmed in our model, as in Ref.~\refcite{Chen:2017xat}  the $\Xi^{(*)}D^{(*)}$ channels were not considered.

\subsection{$\Xi_c$ excited states}
\label{sec:xic}

In the $\Xi_c$ sector we proceed in a similar way as done for the $\Omega_c$ states by revising the results of Ref.~\refcite{Romanets:2012hm}. There, nine states were obtained with masses below or close to the experimental $\Xi_c(2790)$/$\Xi_c(2815)$ states, while being far below in mass with respect to $\Xi_c(2930)$/$\Xi_c(2970)$.  Now we identify our $\Xi_c$ states in the cutoff scheme and assess the dependence of our predictions on the value of the cutoff. 

In this manner, we arrive to several  conclusions, extensively discussed in Ref.~\refcite{Nieves:2019jhp}. We find that the experimental $\Xi_c(2790)$, $\Xi_c(2815)$, $\Xi_c(2930)$ and $\Xi_c(2970)$ states can be described as molecules. In particular, we determine that the $\Xi_c(2790)$ state has a large molecular $\Lambda_c \bar K$ component, with a dominant $j_{\it ldof}^\pi=0^-$ configuration for the light degrees of freedom (ldof). We find that $\Xi_c(2790)$ could be  the sibling of the narrow $\Lambda_c(2595)$ in the double pole description of the $\Lambda_c(2595)$. We determine that  the $3/2^-$ $\Lambda_c(2625)$ and $\Xi_c(2815)$ states cannot be SU(3) siblings, in particular given the fact that  $\Lambda_c(2625)$ is probably a constituent three quark state \cite{Yoshida:2015tia,Nieves:2019nol}. Moreover, we predict the existence of other $\Xi_c-$states, not experimentally detected yet, being one of them the sibling of the wide  $\Lambda_c(2595)$. Interestingly, the recently discovered $\Xi_c(2930)$ and $\Xi_c(2970)$ are  HQSS partners. The $\Xi_c(2930)$ would belong to a SU(3) sextet, where there is also a $\Omega_c$ state. This state would correspond to either the experimental $\Omega_c(3090)$ or the  $\Omega_c(3119)$, given their  decay channels. Assuming the equal spacing rule,  we determine a $J=1/2^-$ $\Sigma_c$ state around 2800 MeV that will complete the sextet and fits clearly with the experimental $\Sigma_c(2800)$  \cite{Tanabashi:2018oca}.

The description of the excited $\Xi_c$ states determined by the local hidden gauge model of Ref.~\refcite{Yu:2018yxl} is again different to ours due to the distinct RS and the interactions involving $D$ and $D^*$ and light vector mesons with baryons.

\section{Conclusions}
\label{sec:conclusions}

In view of the recent  observation of excited $\Omega_c$ and $\Xi_c$ states by the LHCb and Belle Collaborations, we have explored their possible interpretation as molecular states. We have used a coupled-channel unitarized model  based on a ${\rm SU(6)}_{\rm lsf}\times$HQSS-extended WT baryon-meson interaction, revising the  predictions of Ref.~\refcite{Romanets:2012hm}. The present work is a summary of Refs.~(\refcite{Nieves:2017jjx,Nieves:2019jhp}).
 
With regards to the five $\Omega_c$ excited states, we conclude that  (probably) at least three of the states observed by LHCb have spin-parity $J=1/2^-$ and $J=3/2^-$. Also, we find that 
 $\Xi_c(2790)$, $\Xi_c(2815)$, $\Xi_c(2930)$ and $\Xi_c(2970)$ states can be described as molecules. Interestingly,  $\Xi_c(2930)$ and $\Xi_c(2970)$ would be HQSS partners, and   $\Xi_c(2930)$ would be part of a SU(3) sextet containing either the $\Omega_c(3090)$ or the $\Omega_c(3119)$, and that would be completed by the experimental $\Sigma_c(2800)$.

\section*{Acknowledgments}
R. P. Pavao acknowledges support from the Generalitat Valenciana in the program GRISOLIAP/2016/071. This
research has been supported by the Deutsche Forschungsgemeinschaft under
Project Nr. 383452331 (Heisenberg Programme) and Project Nr. 411563442;  Spanish Ministerio de Econom\'ia y
Competitividad and the European Regional Development Fund, under
contracts FIS2017-84038-C2-1-P, FPA2013-43425-P,
FPA2016-81114-P and SEV-2014-0398;  the THOR COST Action CA15213; and by the EU STRONG-2020 project under the program
H2020-INFRAIA-2018-1, grant agreement no. 824093.

\bibliographystyle{ws-procs9x6} % for numbered citation & references
\bibliography{ref.bib}

\end{document}